\documentclass[showpacs,preprintnumbers,superscriptaddress]{revtex4}
\usepackage{}

\usepackage{amsmath,amssymb,graphicx,bm}

\begin{document}
%\begin{CJK*}
%\preprint{APS/123-QED}

\title{Dynamics of generalized tachyon field}

\author{Rongjia Yang}
\email{yangrongjia@tsinghua.org.cn}
 \affiliation{College of Physical Science and Technology, Hebei University, Baoding 071002, China}
 \affiliation{Department of Physics, Tsinghua University, Beijing 100084, China}

\author{Jingzhao Qi}
 \affiliation{College of Physical Science and Technology, Hebei University, Baoding 071002, China}

%\date{\today}

\begin{abstract}
We investigate the dynamics of generalized tachyon field in FRW spacetime. We obtain the autonomous dynamical system for the general case. Because the general autonomous dynamical system cannot be solved analytically, we discuss two cases in detail: $\beta=1$ and $\beta=2$. We find the critical points and study their stability. At these critical points, we also consider the stability of the generalized tachyon field, which is as important as the stability of critical points. The possible final states of the universe are discussed.


\end{abstract}

\maketitle
%\end{CJK*}

\section{Introduction}
An unknown energy component, dubbed dark energy, is usually proposed to explain the accelerated expansion. The simplest and most attractive candidate is the
cosmological constant model ($\Lambda$CDM) with a constant
equation of state (EoS) parameter $w=-1$. This model is consistent
with the current astronomical observations, but is not well regarded because of the cosmological constant problem \cite{weinberg} as well as age problem \cite{Yang2010}.
It is thus natural to pursue alternative possibilities to explain
the mystery of dark energy. Over the past numerous dark energy models have been proposed, such as
quintessence, phantom, k-essence, tachyon, etc. These scalar field models can be extended to a more general model with Lagrangian: $\mathcal{L}_{\phi}=f(\phi)F(X)-V(\phi)$, with the kinetic energy $X\equiv \frac{1}{2}\partial_{\mu}\phi\partial^{\mu}\phi$ \cite{Carroll2003,Malquarti2003}. This general Lagrangian has received much attention.
For some special cases of this Lagrangian, theoretical and observational constraints had been considered in \cite{Yang2009,Yang2008,Yang2008a,yang10,yang2011}; phase-space analysis had been investigated in \cite{yang2011a,yang2012}. The geometrical diagnostic had been used to discriminate a special case of this Lagrangian from $\Lambda$CDM \cite{Gao2010}. Recently, the dynamical system and bounce solutions of $F(X)-V(\phi)$ theories were discussed in \cite{Santiago}.
When this Lagrangian takes the form of generalized tachyon, the EoS parameter and the speed of sound can take the same values of generalized quintessence, the two models of dark energy are indistinguishable from the evolution of background as well as from the evolution of perturbations from a Friedmann-Robertson-Walker (FRW) metric \cite{Unnikrishnan2008}. Though it had been studied intensively in \cite{Unnikrishnan2008}, generalized tachyon field model is still worth investigating in a systematic way to inspect the possible final state of the universe.

The aim of this paper is to analyze the possible cosmological
behavior of the generalized tachyon field in FRW spacetime.
We are interesting in investigating the possible late-time solutions which can be obtained by performing a phase-space and stability analysis. In these solutions
we calculate various observable quantities, such as the density of the
dark energy and the EoS parameter. As we shall see, indeed the generalized tachyon cosmology can be consistent with
observations.

This paper is organized as follows: in the following section, we review the model of generalized tachyon field. In Sec. III, we consider the dynamics of generalized tachyon field. In Sec. IV, we discuss the stability of both critical points and the model. Finally, we shall close with a few concluding remarks in Sec. IV.

\section{Generalized tachyon cosmology}
Scalar fields, such as quintessence, phantom, k-essence, tachyon, can act as sources of dark energy. In general the Lagrangian for such scalar fields can be expressed
as \cite{Carroll2003,Malquarti2003}
\begin{eqnarray}
\label{action}
\mathcal{L}_{\phi}=f(\phi)F(X)-V(\phi),
\end{eqnarray}
where $f(\phi)$ and $V(\phi)$ are functions in terms of a scalar field $\phi$. We assume a flat and homogeneous FRW spacetime and work in units $8\pi G=c=1$. Here we consider the generalized tachyon field which had been studied in reference \cite{Unnikrishnan2008}
\begin{eqnarray}
\label{p}
\mathcal{L}_{\phi}=-V(\phi)\left[1-2X\right]^\beta.
\end{eqnarray}
For $\beta=1/2$, Lagrangian (\ref{p}) is the usual Dirac-Born-Infeld form of the Lagrangian (called tachyon) discussed in \cite{Bagla2003,Trodden2003} (see, for a review \cite{Bamba2012}). Here we do not consider this case. For arbitrary $\beta$, to make sense of Lagrangian (\ref{p}), we must have a constraint on $X$: $X\leq 1/2$. For a constant potential $V_0$, a model of generalized tachyon field have been discussed in Refs. \cite{Yang2008,Chimento}. The pressure of generalized tachyon field is $p_\phi=\mathcal{L}_{\phi}$, and the energy density takes the form
\begin{eqnarray}
\rho_\phi = V (\phi)  \left[1+2 \left( 2\beta-1 \right)X \right]  \left[ 1- 2X \right] ^{\beta-1},
\end{eqnarray}
The corresponding EoS parameter and the effective sound speed are given by
\begin{eqnarray}
\label{w}w_\phi&=&\frac{p_\phi}{\rho_\phi}=\frac{2X-1}{1+2(2\beta-1)X}, \\
\label{c}c^{2}_{\rm s}&=&\frac{\partial p_\phi/\partial
X}{\partial\rho_\phi /\partial X}=\frac{2X-1}{4\beta X-2X-1},
\end{eqnarray}
The definition of the sound speed comes from the equation describing the evolution of linear
adiabatic perturbations in a scalar field dominated universe \cite{Garriga1999}. In a flat and homogeneous FRW space-time, the equation for the scalar field takes the form
\begin{eqnarray}
\label{field}
\frac{d}{dt}\left[\frac{\partial \mathcal{L}_{\phi}}{\partial
X}\dot{\phi}\right]+3H\frac{\partial\mathcal{L}_{\phi}}{\partial
X}\dot{\phi}+\frac{\partial\mathcal{L}_{\phi}}{\partial
\phi}=0,
\end{eqnarray}
where $H=\dot{a}/a$ is the Hubble parameter related to the Friedmann equations,
\begin{eqnarray}
\label{f1}
H^2=\frac{1}{3}(\rho_{\rm m}+\rho_{\phi}),\\
\label{f2}
\dot{H}=-\frac{1}{2}(\rho_{\rm m}+\rho_{\phi}+p_{\phi}).
\end{eqnarray}
Here we neglect baryonic matter and radiation for simplicity. One can straightforwardly include them when necessary. To perform the phase-space and stability analysis, we will transform Eqs. (\ref{f1}) and (\ref{f2}) into an autonomous dynamical system in next section.

\section{The basic equations and the critical points}
In order to transform the cosmological equations into an autonomous dynamical system, it is convenient to introduce auxiliary variables:
\begin{eqnarray}
x=\dot{\phi},~~~~y=\frac{\sqrt{V(\phi)}}{\sqrt{3}H}.
\end{eqnarray}
Using these variables, we straightforwardly obtain the density parameter of dark energy,
\begin{eqnarray}
\label{constraint}
\Omega_\phi=\frac{\rho_\phi}{3H^2}=y^2[1+(2\beta-1)x^2](1-x^2)^{\beta-1}.
\end{eqnarray}
Because $0\leq \Omega_\phi \leq 1$, this gives constraints on $x$ and $y$. In general, when the auxiliary variable $x$ and $y$ can take infinite values,
it is necessary to analyze the dynamics at infinity by using Poincar\'{e} Projection method \cite{Carloni, Carloni2006}. However, because of the constraints (\ref{constraint}) it is not necessary to analyze the dynamics at infinity in the case we discussed here. The EoS, the sound speed of generalized tachyon field, and the total EoS are reformulated as, respectively,
\begin{eqnarray}
w_\phi&=&\frac{x^2-1}{1+(2\beta-1)x^2},\\
\label{sound}
c^{2}_{\rm s}&=&\frac{x^2-1}{(2\beta-1)x^2-1},\\
w_{\rm t}&=&\frac{p_\phi}{\rho_\phi+\rho_{\rm m}}=-y^2(1-x^2)^{\beta}.
\end{eqnarray}
Eqs. (\ref{field}), (\ref{f1}) and (\ref{f2}) give a self-autonomous system in terms of the auxiliary variables $x$ and $y$:
\begin{eqnarray}
\label{aut}
x' &=&-\frac{1}{2}{\frac { \left( - \sqrt {3} \lambda y x^{2}+2\sqrt {3}\lambda \beta y x^{2}
-6\beta x+\sqrt {3}\lambda y \right)  \left( -1+{x}^{2}
 \right) }{\beta \left( -1-x^{2}+2\beta x^{2} \right) }}\\
\label{autd}
y' &=& -\frac{1}{2}y \left[ \sqrt {3}\lambda xy-3+3\, \left( 1-x^{2} \right) ^{
\beta} y^{2} \right]
\end{eqnarray}
where $\lambda \equiv -V_\phi/V^{\frac{3}{2}}$ and the prime denotes a derivative with respect to the logarithm of the scale factor, $N\equiv \ln a$. Here we only consider the case where $\lambda$ is a constant, that is to say $V(\phi)\propto \phi^{-2}$. So in this case, equations (\ref{aut}) and (\ref{autd}) form an autonomous dynamical system. This self-autonomous system are valid in the whole phase-space, not only at the critical points. The critical points $(x_{\rm c},y_{\rm c})$ of the autonomous system are obtained by setting the left-hand sides of
the equations to zero, namely let $\textbf{X}'=(x', y')^T=0$. In order to determine the stability properties of these critical points we expand $\textbf{X}$ around $\textbf{X}_{\rm c}$, setting $\textbf{X}=\textbf{X}_{\rm c}+\textbf{U}$ with the perturbation of the variables $\textbf{U}$ (see, for example, Refs. \cite{Copeland1998,yang2011a,Capozziello,Leon2009}). Thus, up to the first order we acquire $\textbf{U}'=\textbf{M}\cdot \textbf{U}$, where the matrix $\textbf{M}$ contains the coefficients of the perturbation equations. Thus, for each critical point, the eigenvalues of $M$ determine its type and stability. The conditions for the stability of the critical points are Tr $\textbf{M}<0$ and $\det \textbf{M}>0$.

For hyperbolic critical
points, all the eigenvalues have real parts different from zero, one can easily extract their type: source (unstable) for positive
real parts, saddle for real parts of different sign, and sink
(stable) for negative real parts. However, if at least one
eigenvalue has a zero real part (non-hyperbolic critical point), one is not able to obtain conclusive information about the
stability from linearization and needs to resort to other tools
like Normal Forms calculations  \cite{arrowsmith,wiggins}, or
numerical experimentation.

\subsection{The case for $\beta=1$}
\begin{figure}
\includegraphics[width=10cm]{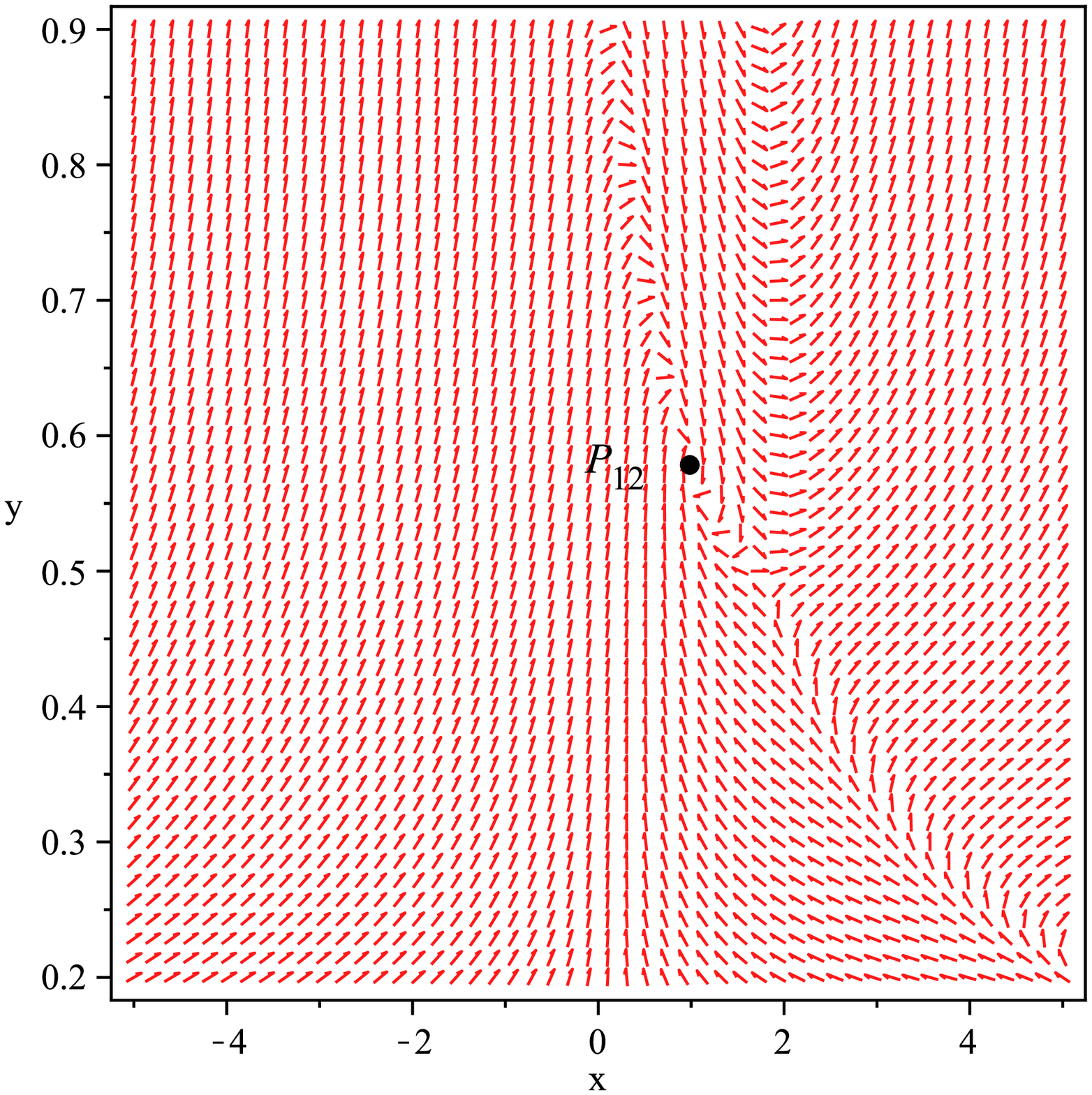}
\caption{Phase-space for generalized tachyon field cosmology, with the choice $\lambda=3$ for critical point $P_{12}$. \label{Fig2}}
\end{figure}

\begin{figure}
\includegraphics[width=10cm]{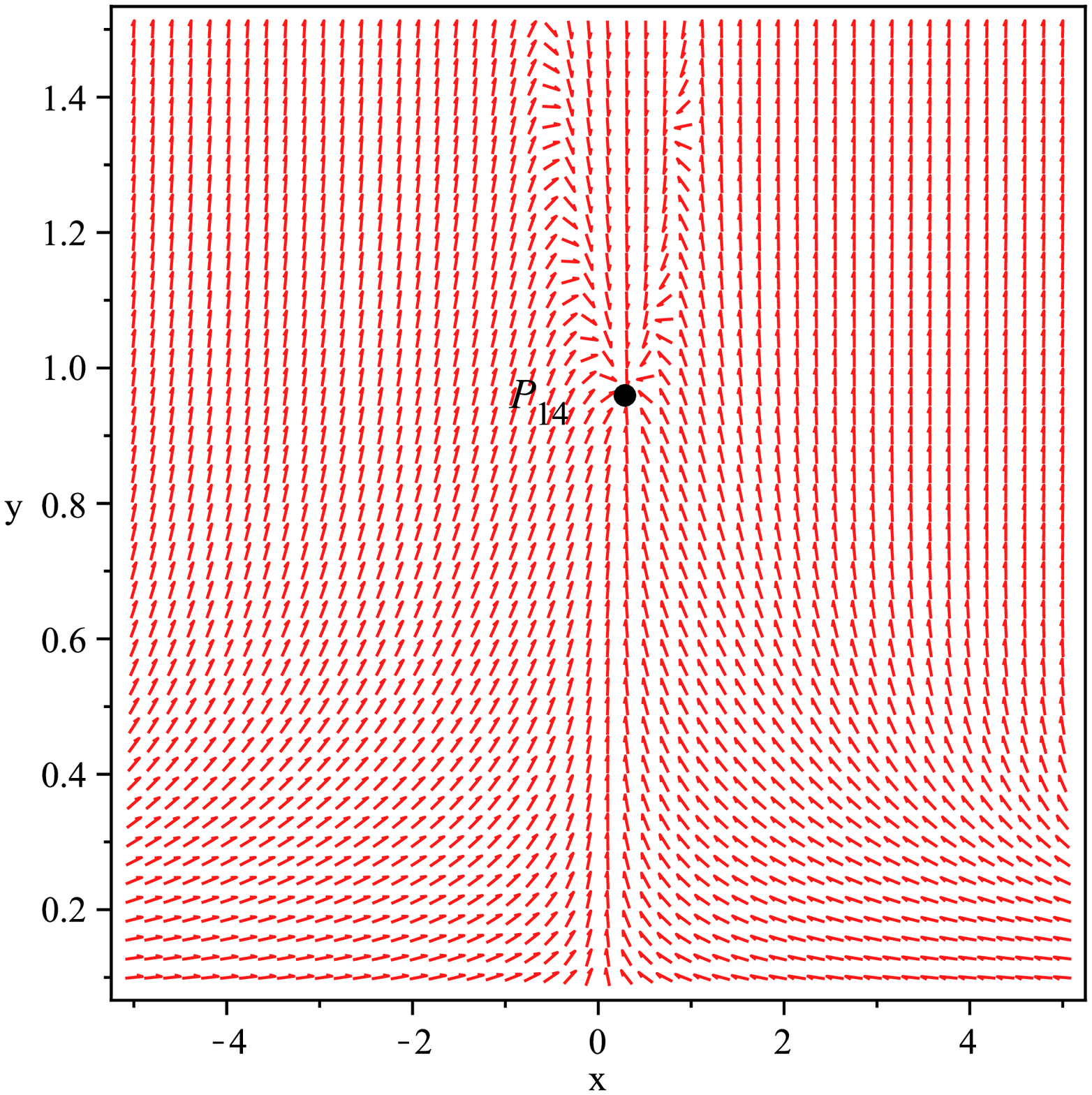}
\caption{Phase-space for generalized tachyon field cosmology, with the choice $\lambda=1$ for critical point $P_{14}$. \label{Fig4}}
\end{figure}

For a arbitrary $\beta$, equations (\ref{aut}) and (\ref{autd}) cannot be analytically resolved. So we investigate two cases for a certain value of $\beta$. One case is $\beta=1$, the other case is $\beta=2$. These two cases are not only simpler, but also interesting in physics, as we will see below. Firstly, we consider the case for $\beta=1$. Then the Lagrangian (\ref{p}) is
\begin{eqnarray}
\label{p1}
\mathcal{L}_{\phi}=p_\phi=V(\phi)\dot{\phi}^2-V(\phi).
\end{eqnarray}
This Lagrangian generalized the quintessence dark energy model and has not been discussed before. So it is worth to investigate this model in detail. Equations (\ref{aut}) and (\ref{autd}) take the form
\begin{eqnarray}
\label{aut1}
x' &=& \frac{1}{2} \sqrt {3}\lambda y x^{2}+\frac{1}{2} \sqrt {3}\lambda y-3\,x,\\
y' &=& \frac{1}{2}\,y \left( -\sqrt {3}\lambda\,xy+3-3\,{y}^{2}+3\,{y}^{2}{x}^{2}
 \right).
\end{eqnarray}
After some algebraic calculus, we obtain the critical points as shown in Table \ref{crit}. The $2\times2$ matrix ${\bf {M}}$ of the linearized perturbation equations is
\[\textbf{M}= \left[ \begin{array}{ccc}
 -3+\sqrt{3}\lambda x_{\rm c}y_{\rm c} & & \frac{\sqrt{3}}{2} \lambda (1+x^2_{\rm c}) \\
 -\sqrt{3}y^2_{\rm c} (\sqrt{3}y_{\rm c}x_{\rm c}-\frac{\lambda}{2})  & & -\sqrt{3}\lambda x_{\rm c}y_{\rm c}+\frac{9}{2} x^2_{\rm c}y^2_{\rm c}-\frac{9}{2} y^2_{\rm c}+\frac{3}{2}
\end{array} \right],\]
According to the conditions for the stability of the critical points, we obtain the ranges of $\lambda$
to make the critical points stable, as shown in table \ref{crit} in which we also present the necessary conditions for their existence,
as well as the corresponding cosmological parameters, $c^{2}_{\rm s}$, $\Omega_{\phi}$, $w_{\phi}$, and $w_{\rm t}$. With these cosmological parameters, we can investigate the final state of the universe and discuss whether there exists acceleration phase or not. From Table \ref{crit}, we can see that points $P_{11}$ and $P_{15}$ are unstable for all $\lambda$; points $P_{12}$, $P_{13}$, and $P_{14}$ are stable for a certain range of $\lambda$. In order to have a visual understanding of the behavior of the field near the critical points, we plot critical point $P_{12}$ for $\lambda=3$ in Fig. \ref{Fig2} and $P_{14}$ for $\lambda=1$ in Fig. \ref{Fig4}.

%%%%%%%%%%%%%%%%%%%%%
\begin{table*}
\begin{center}
\begin{tabular}{|c|c|c|c|c|c|c|c|c|}
  \hline
  Cr. P & \{$x_{\rm c}, y_{\rm c}$\} & Existence & Stable for & $c^{2}_{\rm s}$ & $\Omega_{\phi}$ & $w_{\phi}$ & $w_{\rm t}$ & Acceleration \\\hline
  $P_{11}$ & $\{0, 0\}$ & all $\lambda$ & none & 1 & 0 & -1 & 0 & none \\\hline
  $P_{12}$ & $\{1, \frac{\sqrt{3}}{\lambda}\}$ & $\sqrt{6} <\lambda < \infty$ & $\sqrt{6} <\lambda < \infty$ & 1 & $\frac{6}{\lambda^2}$ & 0 & 0 & none \\\hline
  $P_{13}$ & $\{-1, -\frac{\sqrt{3}}{\lambda}\}$ & $-\infty <\lambda <\sqrt{6}$ & $-\infty <\lambda <\sqrt{6}$ & 1 & $\frac{6}{\lambda^2}$ & 0 & 0 & none \\\hline
  $P_{14}$ & $\{\frac{\lambda}{\lambda_0}, \frac{\sqrt{3}}{6}\lambda_0\}$ & $-2\sqrt{3} < \lambda < 2\sqrt{3}$ & $-\sqrt{6} < \lambda < \sqrt{6}$ & 1 & 1 & $-1+\frac{\lambda^2}{6}$ & $-1+\frac{\lambda^2}{6}$ & $-2<\lambda<2$ \\\hline
  $P_{15}$ & $\{-\frac{\lambda}{\lambda_0}, -\frac{\sqrt{3}}{6}\lambda_0\}$ & none & none & 1 & 1 & $-1+\frac{\lambda^2}{6}$ & $-1+\frac{\lambda^2}{6}$ & none \\\hline
\end{tabular}
\end{center}
\caption{\label{crit} The cosmological parameters and the behavior of the critical points with $\lambda_0=\sqrt{12-\lambda^2}$.}
\end{table*}
%%%%%%%%%%%%%%%%%%%%%%%%

%%%%%%%%%%%%%%%%%%%%%%%%%%%%%%%%%%%%%%%%%%%%%%
%%%%%%%%%%%%%%%%%%%%%%%%%%%%%%%%%%%%%%%%%%%%%%%%%

%%%%%%%%%%%%%%%%%%%%%%%%%%%%%%%%%%%%%%%%%%%%%%
%%%%%%%%%%%%%%%%%%%%%%%%%%%%%%%%%%%%%%%%%%%%%%%%%

\subsection{The case for $\beta=2$}
\begin{figure}
\includegraphics[width=10cm]{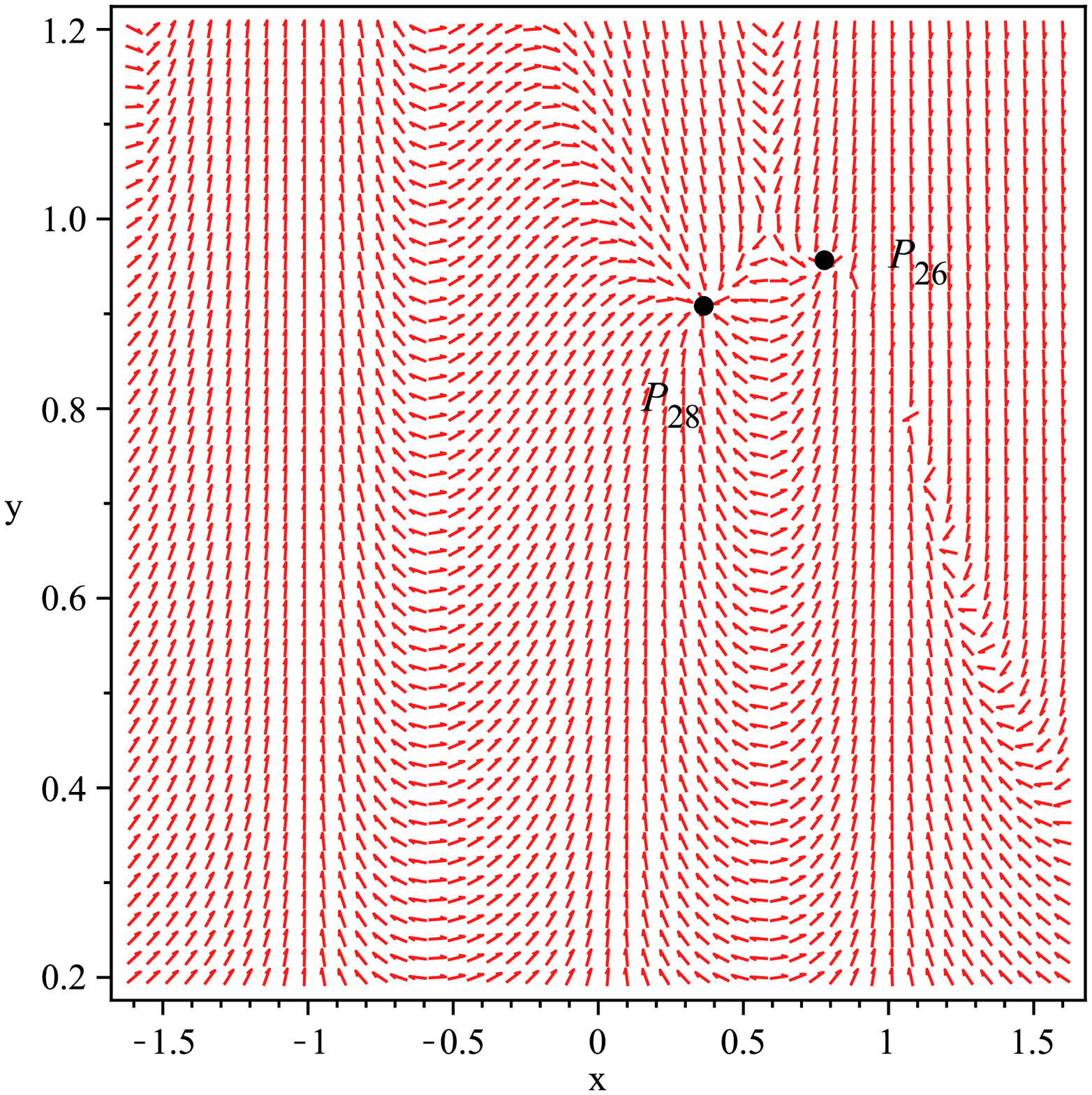}
\caption{Phase-space for generalized tachyon field cosmology, with the choice $\lambda=2$ for critical points $P_{26}$ and $P_{28}$. \label{Fig26}}
\end{figure}

\begin{figure}
\includegraphics[width=10cm]{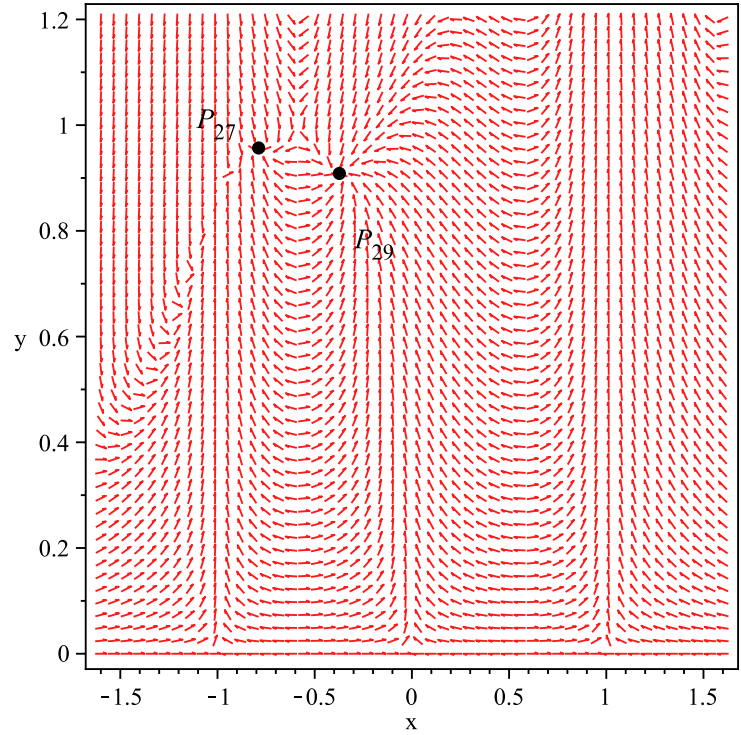}
\caption{Phase-space for generalized tachyon field cosmology, with the choice $\lambda=-2$ for critical points $P_{27}$ and $P_{29}$. \label{Fig27}}
\end{figure}

Secondly, we consider the case for $\beta=2$. The Lagrangian (\ref{p}) changes into
\begin{eqnarray}
\label{p2}
\mathcal{L}_{\phi}=p_\phi=-V(\phi)\dot{\phi}^4+2V(\phi)\dot{\phi}^2-V(\phi).
\end{eqnarray}
This Lagrangian has also not been discussed before and generalized the power law k-essence model, $p_\phi=\frac{1}{4}V(\phi)\dot{\phi}^4-\frac{1}{2}V(\phi)\dot{\phi}^2$, as discussed in \cite{Gao2010}. So it is also worth to consider this case in detail. Equations (\ref{aut}) and (\ref{autd}) take the form
\begin{eqnarray}
\label{aut2}
x' &=& \frac{1}{4}{\frac {-2\sqrt {3}\lambda \,y\,{x}^{2}+3\sqrt {3}\lambda\,y\,{x}^
{4}-\sqrt {3}\lambda y+12\,x-12\,{x}^{3}}{-1+3\,{x}^{2}}}
\\
y' &=& -\frac{1}{2}y \left( \sqrt {3}\lambda\,xy-3+3\,{y}^{2}-6\,{y}^{2}{x}^{2}+3\,
{y}^{2}{x}^{4} \right).
\end{eqnarray}
According to these two equations, we obtain the critical points as shown in Table \ref{crit1}. The $2\times2$ matrix ${\bf {M}}$ of the linearized perturbation equations is
\[\textbf{M}= \left[ \begin{array}{ccc}
 -3+\sqrt{3}\lambda x_{\rm c}y_{\rm c} & & \frac{\sqrt{3}\lambda }{4} \frac{3x^4_{\rm c}-2x^2_{\rm c}-1}{3x^2_{\rm c}-1} \\
-\frac{\sqrt{3}}{2}\lambda y^2_{\rm c}+6 y^3_{\rm c} x_{\rm c}-6 y^3_{\rm c} x^3_{\rm c} & & -\sqrt{3}\lambda x_{\rm c}y_{\rm c}-\frac{9}{2} x^4_{\rm c}y^2_{\rm c}+9 x^2_{\rm c}y^2_{\rm c}-\frac{9}{2} y^2_{\rm c}+\frac{3}{2}
\end{array} \right],\]
with which and the conditions for the stability of the critical points, we obtain the values of $\lambda$ to make critical points stable. From Table \ref{crit1}, we find that points $P_{21}$, $P_{22}$, $P_{23}$, $P_{24}$, and $P_{25}$ are unstable for all $\lambda$; points $P_{26}$, $P_{27}$, $P_{28}$ and $P_{29}$ are stable for a certain range of $\lambda$. The conditions for existence and the corresponding cosmological parameters are also presented in Table \ref{crit1}. We plot critical points $P_{26}$ and $P_{28}$ for $\lambda=2$ in Fig. \ref{Fig26}. Critical points $P_{27}$ and $P_{29}$ for $\lambda=-2$ are also shown in Fig. \ref{Fig27}.

%%%%%%%%%%%%%%%%%%%%%
\begin{table*}
\begin{center}
\begin{tabular}{|c|c|c|c|c|c|c|c|c|}
  \hline
  Cr. P & \{$x_{\rm c}, y_{\rm c}$\} & Existence & Stable for & $c^{2}_{\rm s}$ & $\Omega_{\phi}$ & $w_{\phi}$ & $w_{\rm t}$ & Acceleration \\\hline
  $P_{21}$ & $\{0, 0\}$                          & all $\lambda$         & none & 1 & 0 & -1 & 0 & all $\lambda$ \\\hline
  $P_{22}$ & $\{1, 0\}$                          & all $\lambda$         & none & 0 & 0 & 0 & 0 & none \\\hline
  $P_{23}$ & $\{-1, 0\}$                         & all $\lambda$         & none & 0 & 0 & 0 & 0 & none \\\hline
  $P_{24}$ & $\{-1, -\frac{\sqrt{3}}{\lambda}\}$ & $-\infty < \lambda < 0$ & none & 0 & 0 & $0$ & 0 & none \\\hline
  $P_{25}$ & $\{1, \frac{\sqrt{3}}{\lambda}\}$   & $0 < \lambda < +\infty$ & none & 0 & 0 & $0$ & 0 & none \\\hline
  $P_{26}$ & $\{\frac{\sqrt{6\lambda_1}}{24}, \frac{\sqrt{18\lambda_1}(\lambda_2+32)}{768\lambda}\}$ & $0 < \lambda < \frac{4\sqrt{3}}{3}$ & $0 < \lambda < \frac{4\sqrt{3}}{3}$ & $\frac{\lambda_1-96}{3(\lambda_1-32)}$ & 1 & $\frac{\lambda_1-96}{3(\lambda_1+32)}$ & $-\frac{\lambda_2}{96}$ & none \\\hline
  $P_{27}$ & $\{-\frac{\sqrt{6\lambda_1}}{24}, -\frac{\sqrt{18\lambda_1}(\lambda_2+32)}{768\lambda}\}$ & $-\frac{4\sqrt{3}}{3} < \lambda < 0$ & $-\frac{4\sqrt{3}}{3} < \lambda < 0$ & $\frac{\lambda_1-96}{3(\lambda_1-32)}$ & 1 & $\frac{\lambda_1-96}{3(\lambda_1+32)}$ & $-\frac{\lambda_2}{96}$ & none \\\hline
  $P_{28}$ & $\{\frac{\sqrt{6\lambda_2}}{24}, \frac{\sqrt{18\lambda_2}(\lambda_1+32)}{768\lambda}\}$ & $0 < \lambda < \frac{4\sqrt{3}}{3}$ & $0 < \lambda < \frac{4\sqrt{3}}{3}$ & $\frac{\lambda_1+6\lambda^2}{3(32-\lambda_2)}$ & 1 & $-\frac{\lambda_1+6\lambda^2}{3(\lambda_2+32)}$ & $-\frac{\lambda_1}{96}$ & $0 < \lambda < \frac{4\sqrt{3}}{3}$ \\\hline
  $P_{29}$ & $\{-\frac{\sqrt{6\lambda_2}}{24}, -\frac{\sqrt{18\lambda_2}(\lambda_1+32)}{768\lambda}\}$ & $-\frac{4\sqrt{3}}{3} < \lambda < 0$ & $-\frac{4\sqrt{3}}{3} < \lambda < 0$ & $\frac{\lambda_1+6\lambda^2}{3(32-\lambda_2)}$ & 1 & $-\frac{\lambda_1+6\lambda^2}{3(\lambda_2+32)}$ & $-\frac{\lambda_1}{96}$ & $-\frac{4\sqrt{3}}{3} < \lambda < 0 $
  \\\hline
\end{tabular}
\end{center}
\caption{\label{crit1} The cosmological parameters and the behavior of the critical points with $\lambda_1=48-3\lambda^2+\sqrt{9\lambda^4-480\lambda^2+2304}$ and $\lambda_2=48-3\lambda^2-\sqrt{9\lambda^4-480\lambda^2+2304}$.}
\end{table*}
%%%%%%%%%%%%%%%%%%%%%%%%

%%%%%%%%%%%%%%%%%%%%%%%%%%%%%%%%%%%%%%%%%%%%%%
%%%%%%%%%%%%%%%%%%%%%%%%%%%%%%%%%%%%%%%%%%%%%%%%%

%%%%%%%%%%%%%%%%%%%%%%%%%%%%%%%%%%%%%%%%%%%%%%
%%%%%%%%%%%%%%%%%%%%%%%%%%%%%%%%%%%%%%%%%%%%%%%%%

\subsection{Stability of model}
As shown in \cite{Gao2010}, the stability of points is related to the perturbations $\delta x$ and $\delta y$, and depends on the condition: Tr $\textbf{M}<0$ and $\det \textbf{M}>0$; the classical stability of the model is related to the perturbations $\delta \rho$, and depends on the condition: $c^{2}_{\rm s}\geq 0$; and the quantum stability of the model is related to the perturbations $\delta \phi$, and depends on the condition: $p_X+2Xp_{XX} \geq 0$ and $p_X \geq 0$. That is to say, the stability of the points is different from the classical/quantum stability of the model. So in order to study the possible final state of the universe, it is important to investigate not only the stability of the critical points but also the stability of the model. Like in \cite{Gao2010}, we must investigate the (classical and quantum) stability of the model. Firstly, we consider the classical stability. The equation for the canonical quantization variable $v$ describing the collective metric and scalar field perturbations can be written down in the standard way and takes the form in a flat Universe \cite{Garriga1999}
\begin{eqnarray}
v''_k+(c^{2}_{\rm s}k^2-\frac{\Phi''}{\Phi})v_k=0,
\end{eqnarray}
where $\Phi=\frac{a(\rho_\phi+p_\phi)^{1/2}}{c_{\rm s}H}$. The increment of instability is inversely proportional to the wave-length of the perturbations,
and therefore the background models for which $c^{2}_{\rm s}<0$ are violently unstable and do not have any physical significance. So we insist on $c_{\rm s}^2 \geq 0$. Another potentially interesting requirement to consider is $c_{\rm s}^2 \leq 1$, which says that the sound speed should not exceed the speed of light, which suggests violation of causality. Note, however, that this is still an open problem (see e. g. \cite{Babichev2008,Bruneton2007,Kang2007,Bonvin2006,Gorini2008,Ellis2007}).

Combining the condition for classical stability and Eq. (\ref{sound}), we obtain the range in which the model is classically stable: $0\leq X\leq$ min $\{\frac{1}{2}, \frac{1}{4\beta-2}\}$
(max$\{-1, -\frac{1}{2\beta-1}\}\leq x \leq $ min $\{1, \frac{1}{2\beta-1}\} $) for $\beta> 1/2$, or $0\leq X\leq 1/2$ ($-1\leq x \leq 1$) for $\beta< 1/2$. For the case of $\beta=1$, we have $c_{\rm s}^2=1$, meaning the model is classically stable. For the case of $\beta=2$, we have $0\leq c_{\rm s}^2\leq1$, also meaning the model is classically stable.

Secondly, we discuss the quantum stability of the generalized tachyon field. We consider the small fluctuations $\delta \phi$ around a background value $\phi_0$ which is a solution of the equations of motions in Minkowski spacetime: $\phi=\phi_0+\delta \phi$. By expanding $p$ at second order in $\delta \phi$, we find the Hamiltonian fluctuations \cite{Picon,Bamba2012,Piazza}:
\begin{eqnarray}
\label{fluc}
\delta \mathcal {H}=(p_X+2Xp_{XX})\frac{(\delta \dot{\phi})^2}{2}+p_X \frac{(\nabla \delta \phi)^2}{2}-p_{\phi\phi}\frac{(\delta \phi)^2}{2},
\end{eqnarray}
where $p_{\phi\phi}\equiv d^2p/d\phi^2$. The positivity of the first two terms in Eq. (\ref{fluc}) leads to the following conditions for stability
\begin{eqnarray}
p_X+2Xp_{XX} \geq 0,~~~~~~p_X \geq 0.
\end{eqnarray}
Inserting (\ref{p}) into those equations above, we obtain the range in which the model is quantum stable: $0\leq X \leq 1/2$ ($-1\leq x \leq 1$) for $\beta< 1/2$, while $0\leq X \leq 1/(4\beta-2)$ ($-1/(2\beta-1)\leq x \leq 1/(2\beta-1)$) for $\beta> 1/2$. We found the model is quantum stable at all critic points for all $\lambda$ for $\beta=1$. The model is quantum stable at critical points $P_{21}$, $P_{28}$, and $P_{29}$, not quantum stable at critical points $P_{22}$, $P_{23}$, $P_{24}$, and $P_{25}$ for $\beta=2$. In Ref. \cite{Kahya}, another method are used to investigate the quantum stability.

So we conclude that the model is both classically and quantum stable for $0\leq X \leq 1/2$ ($-1\leq x \leq 1$) if $\beta< 1/2$, or for $0\leq X\leq$ min $\{\frac{1}{2}, \frac{1}{4\beta-2}\}$ (max$\{-1, -\frac{1}{2\beta-1}\}\leq x \leq $ min $\{1, \frac{1}{2\beta-1}\} $) if $\beta> 1/2$. For example,
the model will be classically and quantum stable in the range of $-1\leq x \leq 1$ for $\beta=1$ or in the range of $-1/3\leq x \leq 1/3$ for $\beta=2$. With these conditions we can determine whether the model is (classically and quantum) stable or not when variation $x$ takes the corresponding values at critical points: if $x_{\rm c}$ is in the range of $x$ allowed by the conditions of (classical and quantum) stability for model, then the model is (classical and quantum) stable at critical points; if $x_{\rm c}$ is not in the range of $x$ allowed by the conditions of (classical and quantum) stability for model, then the model is not (classical and quantum) stable at critical points.

\subsection{Cosmological implications}
According the phase-space analysis of generalized tachyon cosmology,
we now discuss the corresponding cosmological behavior and the possible final states.

For the case of $\beta=1$, the model is classically or quantum stable at all critical points. But critical points $P_{11}$ and $P_{15}$ are not stable, so they are not relevant from a cosmological point of view. At critical points $P_{12}$ and $P_{13}$, the universe is partly occupied by generalized tachyon field with $\Omega_{\phi}=6/\lambda^2$, but the final state seems like a dark matter dominated era because the generalized tachyon field behaves like dark matter: $w_{\phi}=0$. At critical point $P_{14}$, the universe is dominated by generalized tachyon field with $\Omega_{\phi}=1$ and $w_{\phi}=-1+\lambda^2/6$. When $\lambda\rightarrow \pm \sqrt{6}$, the generalized tachyon field will behave like dark matter; when $\lambda=\pm 2$, the universe expands with constant-speed; while for $-2<\lambda<2$, the universe have accelerating phases at this critical points.

For the case of $\beta=2$, critical points $P_{21}$, $P_{22}$, $P_{23}$, $P_{24}$, and $P_{25}$ are unstable, so they are also not physical interesting. Critical points $P_{26}$ and $P_{27}$ are stable for a certain range of values of $\lambda$, but the model are not classically and quantum stable at these two points.
Critical points $P_{28}$ with $0< \lambda<4\sqrt{2}/3$ and $P_{29}$ with $-4\sqrt{2}/3< \lambda< 0$ are more interesting, because not only the points are stable but also the model are classically and quantum stable. Thus they can be the late-time attractors of the universe. At both critical points $P_{28}$ and $P_{29}$, the universe is completely dominated by generalized tachyon field with $w_{\phi}=-\frac{\lambda_1+6\lambda^2}{3(\lambda_2+32)}$ and presents accelerating phases. When $\lambda\longrightarrow 4\sqrt{2}/3 (-4\sqrt{2}/3)$, point $P_{28}$ ($P_{29}$) results to constant-speed expansion. The final state of the universe dependents on the generalized tachyon field and its potential.

\section{Conclusions and discussions}
We have made a comprehensive phase-space analysis of generalized tachyon cosmology. We have examined whether a
universe governed by generalized tachyon can have late-time solutions compatible with observations.

For a arbitrary $\beta$, the autonomous dynamical system cannot be analytically resolved. So we have considered two cases: $\beta=1$ and $\beta=2$. These two cases have not been discussed before and are interesting in physics. We have obtained the critical points and the conditions for their existence and stability. We have also calculated the values of the corresponding cosmological parameters, $c^{2}_{\rm s}$, $\Omega_{\phi}$, $w_{\phi}$, and $w_{\rm t}$, which are important in generalized tachyon cosmology. We have investigated the classical stability of the model, as well as the quantum stability. We have discussed the behavior of the critical points and have plotted some of them.

As shown in \cite{Gao2010}, the critical points can be divided into three classes: unstable points at which the model is stable, stable points at which the model is stable, stable points at which the model is (classically or quantum) unstable. The case for points unstable but model stable or the case for points stable but model unstable are not relevant from a cosmological point of view. Only stable points at which the model is also (classically and quantum) stable are physically interesting. So for the case of $\beta=1$, only points $P_{12}$, $P_{13}$, and $P_{14}$ are cosmological relevant, at which the expansion of the universe can speed down, speed up, or keep constant speed. For the case of $\beta=2$, only points $P_{28}$ and $P_{29}$ are physically interesting. At those two points, the expansion of the universe can also speed down, speed up, or keep constant speed. The final state of the universe dependents on the generalized tachyon field and its potential.

As we have shown, the stability of critical points does not mean the stability of the model, vice versa. In order to study the possible final state of the universe, it is important to investigate not only the stability of the critical points but also the (classical and quantum) stability of the model. The analysis we performed indicates that generalized tachyon cosmology discussed here can be compatible with observations. Theses results can been taken into account if generalized tachyon dark energy passes successfully observational tests which is the subject of other studies.

\begin{acknowledgments}
This study is supported in part by National Natural Science Foundation of China under Grant No. 11147028, and Hebei Provincial Natural Science Foundation of China under Grant No. A2011201147.
\end{acknowledgments}

\bibliography{apssamp}

\end{document}